\documentclass[11pt,a4paper]{article}

\usepackage{jcappub}
\usepackage{multirow}
\pdfoutput=1
\usepackage[utf8]{inputenc}
\usepackage{float}
\usepackage{hyperref}
\usepackage{amsmath}
\usepackage{graphicx}
\usepackage{dcolumn}
\usepackage{bm}
\usepackage{epsfig}
\usepackage{amssymb,latexsym,mathrsfs}
\usepackage{graphicx}
\usepackage{color}
\definecolor{myred}{RGB}{180,50,28}
\definecolor{myblue}{RGB}{2,50,180}
\definecolor{mygreen}{RGB}{2,150,80}

\hypersetup{colorlinks=true,
            citecolor=blue,
            linkcolor=blue,
            urlcolor=blue}

\newcommand{\be}{\begin{equation}}
\newcommand{\ee}{\end{equation}}
\newcommand{\bea}{\begin{eqnarray}}
\newcommand{\eea}{\end{eqnarray}}

\newcommand{\mpl}{M_{*}^{2}}
\newcommand{\bra}{\alpha_{\textrm{B}}}

\newcommand{\kin}{\alpha_{\textrm{K}}}

\newcommand{\rhoma}{\rho_{m}}
\newcommand{\pma}{p_{m}}

\newcommand{\dd}{{\rm{d}}}
\newcommand{\ii}{{\rm{i}}}

\newcommand{\lcdm}{$\Lambda$CDM}

\newcommand{\CLASS}{\textsc{class}}
\newcommand{\hiclass}{{\tt hi\_class}}
\newcommand{\deltanb}{\delta^{\text{Nb}}}

\newcommand{\deltanbprime}{\delta^{\prime\text{Nb}}}
\newcommand{\deltanbprimeprime}{\delta^{\prime\prime\text{Nb}}}
\newcommand{\deltaN}{\delta^{\text{N}}}

\newcommand{\deltaNprime}{\delta^{\prime\text{N}}}
\newcommand{\deltaNprimeprime}{\delta^{\prime\prime\text{N}}}
\newcommand{\gammanb}{\gamma^{\text{Nb}}}
\newcommand{\HTnb}{H_{\text{T}}^{\text{Nb}}}

\newcommand{\HTnbprime}{H_{\text{T}}^{\prime\text{Nb}}}
\newcommand{\HTnbprimeprime}{H_{\text{T}}^{\prime\prime\text{Nb}}}

\title{Relativistic Corrections to the Growth of Structure in Modified Gravity}

\author[a,b,1]{Guilherme Brando,\note{Corresponding author.}}
\author[b]{Kazuya Koyama}
\author[b]{and David Wands}

\affiliation[a]{PPGCosmo, CCE -- Universidade Federal do Esp\'irito Santo,\\
Avenida Fernando Ferrari
514, 29075-910 Vit\'oria, Esp\'irito Santo, Brazil,}
\affiliation[b]{Institute of Cosmology and Gravitation, University of Portsmouth\\ Dennis Sciama
Building, Portsmouth PO1 3FX, United Kingdom}

\emailAdd{gbrando@cosmo-ufes.org}
\emailAdd{kazuya.koyama@port.ac.uk}
\emailAdd{david.wands@port.ac.uk}
\abstract{We present a method to introduce relativistic corrections including linear dark energy perturbations in Horndeski theory into Newtonian simulations based on the N-body gauge approach. 
We assume that standard matter species (cold dark matter, baryons, photons and neutrinos) are only gravitationally-coupled with the scalar field and we then use the fact that one can include modified gravity effects as an effective dark energy fluid in the total energy-momentum tensor. 
In order to compute the scalar field perturbations, as well as the cosmological background and metric perturbations, we use the Einstein-Boltzmann code \hiclass. 
As an example, we study the impact of relativistic corrections on the matter power spectrum in k-essence, a subclass of Horndeski theory, including the effects of massless and massive neutrinos. 
For massive neutrinos with $\sum m_{\nu} = 0.1$ eV, the corrections due to relativistic species (photons, neutrinos and dark energy) can introduce a maximum deviation of approximately $7\%$ to the power spectrum at $k \sim 10^{-3} \ \textrm{Mpc}^{-1}$ at $z=0$, for a scalar field with sound speed $c_{s}^{2}\sim0.013$ during matter domination epoch. Our formalism makes it possible to test beyond \lcdm \ models probed by upcoming large-scale structure surveys on very large scales. 
}
\begin{document}
\maketitle
\flushbottom

\section{Introduction}
Einstein's General Theory of Relativity (GR) continues to pass many cosmological and astrophysical tests, reinforcing its position as the standard description of gravitational interactions in cosmological models. The concordance cosmological model, \lcdm, is built around GR plus the standard model of particle physics, with two additional components, dark matter and a Cosmological Constant ($\Lambda$), added in order to reproduce the observed structure in the Universe (galaxy rotation curves, over and under dense regions of the galaxy distribution, acoustic peaks in the cosmic microwave background, etc) and the current accelerated expansion. One may, however, seek a more fundamental description of either of these two components. In this case, modified gravity may help account for one or other of these components, or both at the same time.

In the coming years, new large-scale structure surveys, such as the Legacy Survey of Space and Time (LSST)~\cite{vro}, Euclid~\cite{euclid} and the Dark Energy Spectroscopic Instrument (DESI)~\cite{desi}, will deliver more precise data about the nature of dark matter, dark energy and the properties of the late-time acceleration of the Universe. With this new and more accurate data, it is also necessary to properly take into account possible deviations from the \lcdm \ model, which requires that the modelling of beyond \lcdm \ models is also accurate at the percent level. 
Over the past decade, a substantial effort has been made to increase the speed and precision of N-body simulations capable of simulating the linear and non-linear nature of the Universe. One specific goal of these techniques is to correctly introduce relativistic effects coming from neutrinos and photons. Several schemes have been outlined and studied in the literature~\cite{brand1,agarwal,bird,villa,casto,emb,adamek,brand2,viel,banerjee,brand3,ali,liu,dakin1,Thomas,Partmann}. Dark energy and/or modified gravity, beyond the Standard Model, has also recently been the subject of much interest~\cite{farbod1,farbod2,farbod3,dakin}. It is known that components with non-zero pressure, such as a scalar field with non-standard kinetic term, can contribute to deviations in the matter power spectrum of the order of tens of percent on scales $k<(10^{-3}-10^{-2})\textrm{Mpc}^{-1}$~\cite{dakin}.

Even though these very large scales can be studied in linear perturbation theory, N-body simulations are still required to characterise the observed galaxies and create mock galaxy catalogues. In addition, the galaxy number counts receive various relativistic corrections along the line of sight \cite{yoo,bonvin,classgal,renk,lombriser}. These number counts rely on the fully non-linear dark matter densities and the linear formula may not be used reliably. The observed galaxy number counts can be constructed by employing ray tracing through the non-linear particle distribution in an N-body simulation~\cite{borzyszkowski,breton,nboisson2}. It is thus necessary to include the effect of non-zero pressure in an N-body simulation to make it fully consistent with the linear relativistic perturbation theory on large scales. One possibility is to utilise relativistic simulations such as {\it gevolution}~\cite{gevolution}. Recently, {\it $k$-evolution}, a relativistic N-body code based on {\it gevolution}, has been developed, which includes clustering dark energy among its cosmological components \cite{farbod1}.

In this work we will adopt the N-body gauge approach developed in~\cite{Nbody1,Nbody2,Nbody3,Nbody4}, and present a method to introduce the relativistic corrections in Horndeski theories~\cite{horn,deffa1,koba} in a Newtonian N-body simulation. The N-body gauge is characterised by the absence of the volume deformation in the metric. This implies that the density which a Newtonian simulation computes by a naïve counting of particles in a given coordinate volume is the same as the relativistic density. We treat the modification of gravity as an effective dark energy fluid. The validity of this approach follows from the Bianchi identities and the conservation of the energy-momentum tensor of ordinary matter species, since there is no interaction between matter and dark energy. Our method is valid as long as the perturbations of the effective dark energy fluid can be described by the linear theory. For example, in the case of k-essence, the linear approximation is valid if the sound speed is not too small. 

The structure of this paper is as follows. In Section~\ref{sec:Nbodygauge} we will introduce the N-body gauge formalism and apply it to modified gravity models. In Section~\ref{sec:Horn}, we introduce the Horndeski theory and describe this theory as an effective dark energy fluid that only couples gravitationally to other species. In Section~\ref{sec:Kess} we present an example of the computation of relativistic corrections in k-essence, which is a subclass of the Horndeski theory, and summarize the main results found. We present our conclusions in Section~\ref{sec:Concl}.

\section{N-body Gauge}\label{sec:Nbodygauge}

\subsection{N-body gauge}

We describe the following scalar metric perturbations about a homogeneous and isotropic Friedmann--Lema\^itre--Robertson--Walker background cosmology
\begin{subequations}
\begin{align}\label{eq:metric-potentials}
  g_{00} &= -a^2 \left( 1 + 2A \right) \,, \\
  g_{0i} &=  a^2\, \ii \hat k_i B \,, \\
  g_{ij} &= a^2 \left[ \delta_{ij} \left( 1 + 2 H_{\mathrm{L}} \right) + 2 \left( \delta_{ij}/3 - \hat{k}_i \hat{k}_j \right) H_{\mathrm{T}} \right] \,.
\end{align}
\end{subequations}
We use the metric conventions of~\cite{Nbody4}; $A$ is the perturbation of the lapse function, $B$ is a scalar perturbation in the shift, and $H_{\rm L}$ and $H_{\rm T}$ are respectively the trace and trace-free scalar perturbations of the spatial metric.
For simplicity we consider a single Fourier mode with comoving wavevector, $\mathbf{k}$, wavenumber $k \equiv |\mathbf{k}|$ and direction $\hat{k}_i \equiv k_i/k$. 

The energy-momentum tensor of all particle species is given by
\begin{subequations}
\begin{align} 
	T^{0}_{\phantom{0}0} &= 
	-\sum_\alpha (\rho_{\alpha} + \delta \rho_\alpha) =
	- \sum_\alpha  \rho_\alpha \left( 1  + \delta_\alpha \right) \equiv -  \rho \left( 1  + \delta \right) \,, \\
	T_{{\phantom{0}}0}^i &=  \sum_\alpha ( \rho_\alpha + p_\alpha ) \,\ii \hat k^i v_\alpha \equiv ( \rho + p )\, \ii \hat k^i v \,, \\ 
	T^{i}_{\phantom{i}j} &= \sum_\alpha ( p_\alpha+\delta p_\alpha ) \delta^i_j + \frac{3}{2}(\rho_{\alpha}+p_\alpha) \left( \delta_j^i/3 - \hat k^i \hat k_j \right) \sigma_\alpha \\ \nonumber 
  &\equiv ( p+\delta p ) \delta^i_j + \frac{3}{2}(\rho+p) \left( \delta_j^i/3 - \hat k^i \hat k_j \right)  \sigma  \,, \label{Tmunu}
\end{align}
\end{subequations}
where the dummy index $\alpha$ runs over all species, $\delta$ is the density contrast, $\sigma$ is the anisotropic stress following the convention of Ref.~\cite{MandB}, $\rho$ and $p$ are the background density and pressure respectively.

Thus far our perturbation variables are in an arbitrary gauge. 
The definition of the N-body gauge~\cite{Nbody4} is such that:
\begin{itemize}
    \item[(i)]
the temporal slicing is fixed by setting $B^{\textrm{Nb}}=v^{\textrm{Nb}}$, making the constant-time hypersurfaces orthogonal to the $4$--velocity of the total matter and radiation content;
\item[(ii)]
the spatial threading is fixed by setting $H_{\textrm{L}}^{\textrm{Nb}}=0$, so that the physical volume element of a spatial $3$--hypersurface coincides with the coordinate volume element, $\mathrm{d}^{3}x$, i.e., the physical volume is not perturbed by metric deformations. 
\end{itemize}
As out pointed in references~\cite{Nbody1,Nbody2,Nbody3}, the spatial gauge choice is equivalent to requiring that the remaining spatial metric potential, $\HTnb$, is related to the comoving curvature perturbation, $\zeta$, as
\begin{equation}\label{eq:zeta}
    H_{\textrm{T}}^{\textrm{Nb}} = 3 \zeta.
\end{equation}
More generally this condition (\ref{eq:zeta}) can be used to select the spatial threading, independently of the temporal gauge choice. In particular the N-boisson gauge~\cite{nboisson1,nboisson2}, combines the spatial gauge condition (\ref{eq:zeta}) with an alternative time slicing, which coincides with that used in the Poisson gauge, $kB=\dot{H}_T$. At linear scales, the N-body and N-boisson gauge are connected by a temporal gauge transformation and either of the gauges can be used. However, at non-linear scales (where the density has become non-linear), the temporal gauge condition of the N-body gauge leads to large metric perturbations, making the N-boisson gauge more useful at small scales/late times. The present work, as stated in the introduction, is valid at linear level only, in which case either of these gauges can be used. Throughout the rest of this work, we will focus our attention to the N-body gauge. We will come back to this point in the discussion. 

The simplicity of the N-body gauge is that in the absence of relativistic species (photons and neutrinos, or dark energy perturbations) both the matter density and the particle trajectories in the N-body gauge coincide at linear order with those in Newtonian N-body simulations (it is a Newtonian motion gauge~\cite{Nbody3}). Thus to track relativistic corrections to the matter density we only need to solve for the relativistic components which we expect to be well described by linear perturbation theory on sufficiently large scales.

We will assume that energy-momentum conservation holds for individual species. As in Newtonian simulations, we treat baryons as pressureless matter at late times. For the gravitational equations, we use the Einstein equations, $G_{\mu \nu} = 8 \pi G T_{\mu \nu}$ where $T_{\mu \nu}$ includes the contribution from Cold Dark Matter (CDM), baryons, photons, neutrinos and dark energy. We emphasise that this does not mean we assume general relativity. As we describe below, we can also include modified gravity effects as an effective dark energy fluid in $T_{\mu \nu}$.

For pressureless matter, i.e. CDM plus baryons, the evolution equations are given by
\begin{subequations}
\begin{align}
    & \deltanbprime_m +k v_m^{\mathrm{Nb}} = 0, \label{eq:consvnb} \\
    & ( \partial_\tau + {\cal H} ) v_m^{\mathrm{Nb}} = -k\left( \Phi +  \gamma^{\text{Nb}}\right),\label{eq:vdivnb}\,
\end{align}
where $^{\prime}$ is the derivative with respect to the conformal time $\tau$, and $\gammanb$ introduces relativistic corrections to the Euler equation (\ref{eq:vdivnb}), which vanishes in the absence of relativistic species. As shown in~\cite{Thomas}, $\gammanb$ is given by:
\begin{equation}\label{eq:k2gammanb}
    k^{2}\gammanb = -(\partial_{\tau} + \mathcal{H} )\HTnbprime + 12 \pi G a^{2} \left(\rho+p\right)\sigma.
\end{equation}
In the N-body gauge, the Bardeen potential $\Phi$ satisfies the relativistic Poisson equation, but with contributions coming from all species including relativistic species 
\begin{equation}\label{eq:poissonnb}
    k^{2}\Phi = 4 \pi G a^{2} \sum_{\alpha} \delta \rho_{\alpha}^{\textrm{Nb}},
\end{equation}
\end{subequations}
$\alpha=\{\textrm{cdm},\textrm{b},\gamma,\nu,\textrm{DE}\}$.
Quantities with a superscript Nb are computed in the N-body gauge. The $(0i)$ component of the Einstein equations gives 
\begin{equation} \label{eq:0i}
   \HTnbprime =  3 \mathcal{H} A^{\text{Nb}}. 
\end{equation}
From the momentum conservation equation, the lapse function of the N-body gauge metric reads:
\begin{equation}\label{eq:xiNb1}
    A^{\text{Nb}} = \frac{1}{\rho + p} \left[ \left(\rho+p\right)\sigma -\delta p^{\text{Nb}} \right],
\end{equation}
with $\delta p^{\text{Nb}}$ the total pressure perturbation in the N-body gauge. 

When the relativistic species fluid quantities are negligible, $\delta p^{\text{Nb}}=\sigma=0$ (for example, for sufficiently late times), one has $ A^{\text{Nb}}=\HTnbprime=0$, which implies $\gammanb=0$. In this limit, equations (\ref{eq:consvnb}), (\ref{eq:vdivnb}) and (\ref{eq:poissonnb}) coincide with the Newtonian ones:
\begin{subequations}\label{eq:Newton}
\begin{align}  
	 \delta_{m}^{\prime \rm N} + k v_{m}^{\rm N} &= 0 \,,  \label{eq:consvN}\\ 
   \left( \partial_\tau + {\cal H} \right) v_{m}^{\rm N} &= -k \Phi^{\rm N} \,,\label{eq:vdivN}\\
   k^2 \Phi^{\rm N} &= 4\pi G a^2 \rho_{m} \delta_{m}^{\rm N} \,, \label{eq:poissonN}
\end{align} 
\end{subequations}
where the superscript N denotes the perturbations in Newtonian theory. 
Combining equations~(\ref{eq:consvN})--(\ref{eq:poissonN}) then yields the familiar second-order differential equation for the Newtonian density perturbation
\begin{equation}\label{eq:newteqN}
    \deltaNprimeprime_{m} + \mathcal{H}\deltaNprime_{m} - 4 \pi G a^{2} \rho_{m} \deltaN_{m} =  0,
\end{equation}

More generally, in the presence of relativistic species, combining equations~(\ref{eq:consvnb})--(\ref{eq:poissonnb}), we obtain the second-order differential equation for the density perturbation in the N-body gauge~\cite{brand,adamek1}:
\begin{equation}\label{eq:newteqGR}
    \deltanbprimeprime_{m} + \mathcal{H}\deltanbprime_{m} - 4 \pi G a^{2} \rho_{m} \deltanb_{m} =  4\pi G a^{2} \delta \rho_{\text{GR}},
\end{equation}
where
\begin{equation}\label{eq:drhoGR}
    \delta \rho_{\text{GR}} = \delta \rho_{\gamma}^{\text{Nb}} + \delta \rho_{\nu}^{\text{Nb}} + \delta \rho_{\text{DE}}^{\text{Nb}} + \delta \rho_{\text{metric}}^{\text{Nb}},
\end{equation}
and we define
\begin{equation}
    k^2 \gamma = 4 \pi G a^{2} \delta \rho_{\text{metric}} \,.
\end{equation}

The homogeneous solution to (\ref{eq:newteqGR}) coincides with that of the Newtonian equation (\ref{eq:newteqN}) and can be obtained from Newtonian simulations. However the full solution to (\ref{eq:newteqGR}) includes relativistic corrections sourced by the quantities $\delta \rho_{\gamma}^{\text{Nb}}$, $\delta \rho_{\nu}^{\text{Nb}}$, $\delta \rho_{\text{DE}}^{\text{Nb}}$ and $\delta \rho_{\text{metric}}^{\text{Nb}}$ which must be evaluated using a relativistic approach, such as linear Einstein-Boltzmann codes.

To make contact with the quantities commonly evaluated in Einstein-Boltzman codes, we can write the N-body gauge density perturbations in terms of those in the Synchronous or Poisson gauges using the linear gauge transformation
\begin{equation}\label{eq:NbGT}
    \delta \rho_{\alpha}^{\text{Nb}} = \delta \rho_{\alpha}^{\text{S/P}} + 3\mathcal{H}\left(1+w_{\alpha}\right)\rho_{\alpha}\frac{\theta_{\text{tot}}^{\text{S/P}}}{k^{2}}, 
\end{equation}
with $\mathcal{H}$ being the conformal Hubble factor, $\mathcal{H}=a^{\prime}/a$, and $\theta_{\textrm{tot}}$ the total peculiar velocity divergence of all species $\left(\theta =\ii k^{j}v_{j}\right)$. 
$\delta \rho_{\gamma}^{\text{Nb}}$, $\delta \rho_{\nu}^{\text{Nb}}$ and $\delta \rho_{\text{DE}}^{\text{Nb}}$ can thus all be evaluated using Equation (\ref{eq:NbGT}). 

The computation of $\gamma$, given by (\ref{eq:k2gammanb}), requires $\HTnbprime$ and $\HTnbprimeprime$. Using equations (\ref{eq:xiNb1}) and (\ref{eq:0i}), we obtain the equation for $\HTnbprime$:
\begin{equation}\label{eq:HTnbdot}
        \HTnbprime = 3\frac{\mathcal{H}}{\rho + p} \left[ \left(\rho+p\right)\sigma - \delta p^{\text{S/P}} + p^{\prime}\frac{\theta_{\textrm{tot}}^{\text{S/P}}}{k^{2}} \right],
\end{equation}
and its derivative
 \begin{equation}\label{eq:HTnbdotdot}
 \begin{split}
     \HTnbprimeprime &= \left[\frac{\mathcal{H}^{\prime}}{\mathcal{H}} - \frac{1}{\left(\rho + p\right)}\left( \rho^{\prime} + p^{\prime}\right)\right]\HTnbprime \\
     &+ 3\frac{\mathcal{H}}{\rho+p}\left[ \left(\rho^{\prime}+p^{\prime}\right)\sigma + \left(\rho+p\right)\sigma^{\prime} - \delta p^{\prime} + p^{\prime\prime}\frac{\theta_{\textrm{tot}}^{\text{S/P}}}{k^{2}} + p^{\prime}\frac{\theta_{\textrm{tot}}^{\prime \text{S/P}}}{k^{2}} \right],
\end{split}
\end{equation}
where we used a linear gauge transformation to obtain the total pressure perturbation in the N-body gauge
\begin{equation}
    \label{eq:NbGT2}
        \delta p ^{\textrm{Nb}} = \delta p^{\textrm{S/P}} - p^{\prime} \frac{\theta_{\textrm{tot}}^{\textrm{S/P}}}{k^{2}}.
\end{equation}
In this way, we can compute Equation (\ref{eq:k2gammanb}) solely in terms of fluid quantities and their time derivatives in the Synchronous or Poisson gauge. 

\subsection{Modified gravity as a fluid}
In this section, we will describe modified gravity as an effective fluid. The common approach, presented in most modified gravity papers, is to treat the extra degrees of freedom in modified gravity as part of the spacetime dynamics, rather than an exotic new matter component. However, one can always move these extra terms originating from the new degrees of freedom to the right hand side of Einstein's equations~\cite{pace,arjona,gleyzes1,deffa2}. 
\begin{equation}\label{eq:EEfluidDE}
    G_{\mu \nu} = 8 \pi G \left(T_{\mu\nu} + E_{\mu \nu}\right),
\end{equation}
where $G_{\mu\nu}$ is the Einstein Tensor, $T_{\mu\nu}$ the ordinary matter energy-momentum tensor and $E_{\mu\nu}$ is an effective energy-momentum tensor that absorbs any effects due to the modification of gravity. Assuming the conservation of the energy-momentum tensor for matter, $\nabla^{\mu} T_{\mu \nu}=0$, the effective energy-momentum tensor is also conserved, $\nabla^{\mu} E_{\mu \nu}=0$, due to the Bianchi identity, $\nabla^{\mu} G_{\mu \nu}=0$. We can thus treat modified gravity effects as an effective non-interacting dark energy fluid $E_{\mu \nu}= T^{\rm DE}_{\mu \nu}$. This effective fluid approach stems from the known degeneracy between modified gravity theories and some dark energy models that exhibit anisotropic stress and a time-dependent pressure perturbation at the linear perturbation level~\cite{kunz,koyama1,joyce}. Once we make this identification, the derivation of N-body gauge equations in the previous section holds even in modified gravity models as long as we can describe their effects using linear perturbation theory. 

\section{Horndeski Gravity}\label{sec:Horn}
In this section, we apply our method to compute relativistic corrections in Horndeski gravity using the N-body gauge. 

\subsection{Background}
Horndeski's scalar-tensor theory is the most general theory that describes an Ostrogradski-instability free scalar field with second-order equations of motion. Its action is given by
\begin{equation}
S[g_{\mu\nu},\phi]=\int\mathrm{d}^{4}x\,\sqrt{-g}\left[\sum_{i=2}^{5}\frac{1}{8\pi G}{\cal L}_{i}[g_{\mu\nu},\phi]\,+\mathcal{L}_{\text{m}}[g_{\mu\nu},\psi_{M}]\right]\,,\label{eq:actionhorn}
\end{equation}
where the $\mathcal{L}_{i}$ terms are:
\begin{subequations}
\begin{eqnarray}
{\cal L}_{2} & = & G_{2}(\phi,\,X)\,,\label{eq:L2}\\
{\cal L}_{3} & = & -G_{3}(\phi,\,X)\Box\phi\,,\label{eq:L3}\\
{\cal L}_{4} & = & G_{4}(\phi,\,X)R+G_{4X}(\phi,\,X)\left[\left(\Box\phi\right)^{2}-\phi_{;\mu\nu}\phi^{;\mu\nu}\right]\,,\label{eq:L4}\\
{\cal L}_{5} & = & G_{5}(\phi,\,X)G_{\mu\nu}\phi^{;\mu\nu}-\frac{1}{6}G_{5X}(\phi,\,X)\left[\left(\Box\phi\right)^{3}+2{\phi_{;\mu}}^{\nu}{\phi_{;\nu}}^{\alpha}{\phi_{;\alpha}}^{\mu}-3\phi_{;\mu\nu}\phi^{;\mu\nu}\Box\phi\right] \,. \label{eq:L5}
\end{eqnarray}
\end{subequations}
$X=-\frac{1}{2}\partial_{\mu}\phi\partial^{\mu}\phi$ is the canonical kinetic term of the scalar field, and $\psi_{M}$ represents the matter fields. To obtain the equations of motion for a homogeneous and isotropic background $\mathrm{d}s^{2}=a(\tau)^{2}\left[-N(\tau)\mathrm{d}\tau^{2}+\mathrm{d}\mathbf{x}^{2}\right]$, one varies (\ref{eq:actionhorn}) with respect to the lapse $N(\tau)$ and scale factor $a(\tau)$, where $\tau$ is the conformal time. The equations of motion (setting $N=1$) become~\cite{hiclass1,defelice}:
\begin{subequations}
\begin{align}
    H^{2} &= \frac{8 \pi G}{3} \left( \sum_{i}\rho_{i} + \rho_{DE} \right) \label{eq:H2}\\
    H^{\prime} &= - 4 \pi G a \left[ \sum_{i}\left(\rho_{i} + p_{i}\right) + \rho_{DE} + p_{DE}\right]\label{eq:Hprime}
\end{align}
\end{subequations}
where
\begin{subequations}
\begin{align}
\frac{8 \pi G}{3}\mathcal{\rho_{\text{DE}}}\equiv & -\frac{1}{3}G_{2}+\frac{2}{3}X\left(G_{2X}-G_{3\phi}\right)-\frac{2H^{3}\phi^{\prime}X}{3a}\left(7G_{5X}+4XG_{5XX}\right)\label{eq:rhoDE}\\
 & +H^{2}\left[1-\left(1-\alpha_{\textrm{B}}\right)M_{*}^{2}-4X\left(G_{4X}-G_{5\phi}\right)-4X^{2}\left(2G_{4XX}-G_{5\phi X}\right)\right]\nonumber \\
\frac{8 \pi G}{3}p_{\text{DE}}\equiv & \frac{1}{3}G_{2}-\frac{2}{3}X\left(G_{3\phi}-2G_{4\phi\phi}\right)+\frac{4H\phi^{\prime}}{3a}\left(G_{4\phi}-2XG_{4\phi X}+XG_{5\phi\phi}\right)\label{eq:pDE}\\
 & -\frac{\left(\phi^{\prime\prime}-aH\phi^{\prime}\right)}{3\phi^{\prime}a}HM_{*}^{2}\alpha_{\textrm{B}} -\frac{4}{3}H^{2}X^{2}G_{5\phi X}-\left(H^{2}+\frac{2H^{\prime}}{3a}\right)\left(1-M_{*}^{2}\right)\nonumber\\
 &+\frac{2H^{3}\phi^{\prime}XG_{5X}}{3a}\,,\nonumber 
\end{align}
\end{subequations}
and the index $i$ runs over the ordinary matter species. $M_{*}^{2}$ and $\alpha_{\textrm{B}}$, as well as the other property functions which will appear below, are defined in Appendix \ref{sec:AppA}.

We can see from the definitions of the dark energy background energy density and pressure that we have a dependency on the time-dependent Planck Mass $M_{*}^{2}$, which introduces an ambiguity in the definition of the effective dark energy energy-momentum tensor. By choosing to write these quantities as in Equations (\ref{eq:rhoDE}) and (\ref{eq:pDE}), we can write the conservation of the dark energy density and equation of state in its usual form:

\begin{subequations}
\begin{align}
    \rho^{\prime}_{DE} &= -3\mathcal{H}\left(\rho_{DE}+p_{DE}\right)\label{eq:rhoprimeDE},\\
    w_{DE} &= \frac{p_{DE}}{\rho_{DE}}\label{eq:wDE}.
\end{align}
\end{subequations}
This is a different definition than the ones found in refs.~\cite{belsaw,gleyzes1}. In these, the conservation of the energy density of the scalar field does not assume its standard form, but rather exhibits an exchange between the matter and the scalar field, as a result of absorbing the time-dependent Planck Mass $M_{*}^{2}$ into the definition of $T_{\mu\nu}$. Our choice of (\ref{eq:rhoprimeDE}) and (\ref{eq:wDE}) follows the definitions of the code \hiclass~\cite{hiclass1,hiclass2}\footnote{\href{http://miguelzuma.github.io/hi_class_public/}{{\tt http://miguelzuma.github.io/hi\_class\_public/}}}. This definition is necessary for the use of the N-body gauge equations derived in the previous section where we assume that dark energy has no interaction with other matter. 

To facilitate the implementation of our work in Einstein-Boltzmann codes, such as \hiclass, we will adopt the notation and conventions of \CLASS~\cite{CLASS1,CLASS2}\footnote{We stress that in \CLASS \ conventions, Equations (\ref{eq:H2}) and (\ref{eq:Hprime}) are rescaled by $8 \pi G/3$, in order to set internal units of all dimensionful quantities in the code to $\textrm{Mpc}^{-1}$. In this way the background equations assume the form:
\begin{align*}
H^{2}= & \sum_{i}\rho_{i}+\rho_{\textrm{DE}}\\
H^{\prime}= & -\frac{3}{2}a\left[\sum_{i}\left(\rho_{i} + p_{i}\right)+\rho_{\textrm{DE}}+p_{\textrm{DE}}\right]\,.
\end{align*}}.

\subsection{Linear perturbations}\label{sec:Hornfluid}
At the linear perturbation level, to fully describe modified gravity as an effective fluid, we need to specify the perturbative fluid quantities: $\delta \rho_{DE}$, $\delta p_{DE}$, $\theta_{DE}$ and $\sigma_{DE}$. In order to do so, we begin by writing the linearly perturbed equations of motion in the synchronous gauge~\cite{hiclass1}\footnote{\hiclass \ is currently only implemented in the synchronous gauge.}. We will follow the conventions of ref.~\cite{MandB} for the synchronous gauge metric
\begin{equation}\label{eq:synchmetric}
    \dd s^{2} = a(\tau)^2\left[-\dd \tau + \left(\delta_{ij}+h_{ij}\right)\dd x^{i}\dd x^{j}\right],
\end{equation}
where
\begin{equation}\label{eq:synchpots}
    h_{ij}(\mathbf{x},\tau) = \int \dd^{3}k \, e^{\ii \mathbf{k}.\mathbf{x}} \, \left[\hat{k}_i \hat{k}_j h(\mathbf{k,\tau}) + \left(\hat{k}_{i}\hat{k}_j - 1/3 \delta_{ij}\right)6\eta(\mathbf{k},\tau)\right].
\end{equation}
By a simple inspection of Equations (\ref{eq:synchpots}) and (\ref{eq:metric-potentials}), one has the following relation between the metric potentials:
\begin{subequations}
\begin{align}
    &h = 6 H_{\textrm{L}}, \\
    &\eta = H_{\textrm{L}} + \frac{H_{\textrm{T}}}{3}.
\end{align}
\end{subequations}
The linearly perturbed equations are:
\begin{subequations}
\begin{itemize}
\item Einstein (0,0)
\begin{align}
h^{\prime}= & \frac{4k^{2}\eta}{aH\left(2-\alpha_{\textrm{B}}\right)}+\frac{6a\delta\rho_{\textrm{m}}}{HM_{*}^{2}\left(2-\alpha_{\textrm{B}}\right)}-2aH\left(\frac{\alpha_{\textrm{K}}+3\alpha_{\textrm{B}}}{2-\alpha_{\textrm{B}}}\right)V_{X}^{\prime}\nonumber \\
 & -2\left[3aH^{\prime}+\left(\frac{\alpha_{\textrm{K}}+3\alpha_{\textrm{B}}}{2-\alpha_{\textrm{B}}}\right)a^{2}H^{2}+\frac{9a^{2}}{M_{*}^{2}}\left(\frac{\rho_{\textrm{m}}+p_{\textrm{m}}}{2-\alpha_{\textrm{B}}}\right)+\frac{\alpha_{\textrm{B}}k^{2}}{2-\alpha_{\textrm{B}}}\right]V_{X}.\label{eq:metric_00}
\end{align}

\item Einstein (0,i)
\begin{align}
\eta^{\prime}= & \frac{3a^{2}\theta_{\textrm{m}}}{2k^{2}M_{*}^{2}}+\frac{aH}{2}\alpha_{\textrm{B}}V_{X}^{\prime}+\left[aH^{\prime}+\frac{a^{2}H^{2}}{2}\alpha_{\textrm{B}}+\frac{3a^{2}}{2M_{*}^{2}}\left(\rho_{\textrm{m}}+p_{\textrm{m}}\right)\right]V_{X}.\label{eq:metric_0i}
\end{align}

\item Einstein (i,j) trace
\begin{align}
Dh^{\prime\prime}= & 2\lambda_{1}k^{2}\eta+2aH\lambda_{3}h^{\prime}-\frac{9a^{2}\alpha_{\textrm{K}}\delta p_{\textrm{m}}}{M_{*}^{2}}+3a^{2}H^{2}\lambda_{4}V_{X}^{\prime}+2a^{3}H^{3}\left[3\lambda_{6}+\frac{\lambda_{5}k^{2}}{a^{2}H^{2}}\right]V_{X}.\label{eq:metric_ii}
\end{align}

\item Einstein (i,j) traceless
\begin{align}
\xi^{\prime}= & \left(1+\alpha_{\textrm{T}}\right)\eta-aH\left(2+\alpha_{\textrm{M}}\right)\xi+aH\left(\alpha_{\textrm{M}}-\alpha_{\textrm{T}}\right)V_{X}-\frac{9a^{2}\sigma_{\textrm{m}}}{2M_{*}^{2}k^{2}}\,,\label{eq:metric_ij}
\end{align}
\end{itemize}
\end{subequations}
where $H$ is the physical-time Hubble factor, related to the conformal one by $\mathcal{H}=aH$, $V_{X}$ is the scalar field perturbation in conformal time:
\begin{equation}
\label{def:VX}
    V_{X} = a \frac{\delta \phi}{\phi^{\prime}}.
\end{equation}
and $\xi=(h^{\prime}+6\eta^{\prime})/2k^{2}$. The functions $\alpha_{i}$ ($i= \textrm{B, M, K, T}$) are the property functions defined in terms of $G_i(\phi, X)$ characterising linear perturbations in Horndeski's theory and $D$ and $\lambda_{i}$ ($i=1,...,8$) are defined in the Appendix \ref{sec:AppA}. The functions $\alpha_i, \lambda_{i}$ and $D$ are determined by the background. By subtracting the contribution from the Einstein tensor, one can write the new terms coming from modified gravity as effective dark energy fluid quantities:
\begin{subequations}
\begin{itemize}
    \item Density perturbation:

\begin{equation}\label{eq:drhoDE}
\begin{split}
    \delta \rho_{DE} &= \delta \rho_{m} \left(-1-\frac{2}{(\bra-2) \mpl}\right) - \frac{2\bra  }{3 a^2 (\bra-2)}k^2 \eta \\
    &+\frac{2H V_{X}}{3a\mpl\left(\bra-2\right)}\Bigg[ a^2 \Big(H^2 \mpl (3 \bra+\kin)+9 (\pma+\rhoma)\Big)\\
    &-3 a \left(\bra-2\right) \mpl H^{\prime}+\bra k^2 \mpl \Bigg] + \frac{3\bra+\kin}{\bra-2}\frac{2H^{2}V_{X}^{\prime}}{3}.
\end{split}
\end{equation}

    \item Velocity divergence:
    
\begin{equation}\label{eq:thetaDE}
\begin{split}
    \left(\rho_{DE}+p_{DE}\right)\theta_{DE} &=  \left[\frac{2 k^2  H'}{3 a}+\frac{1}{3} \alpha_{\textrm{B}} H^2 k^2 +\frac{k^2  (p_{m}+\rho_{m})}{M^{2}_{*}}\right]V_{X}\\
    &+\frac{\alpha_{\textrm{B}} H k^2  }{3 a}V_{X}^{\prime}+\theta_{m} \left(\frac{1}{M^{2}_{*}}-1\right).
\end{split}
\end{equation}

\item Pressure perturbation:
\begin{equation}\label{eq:dpDE}
\begin{split}
    \delta p_{DE} &= \delta p_{m} \left(\frac{\kin}{D \mpl}-1\right) - \frac{2 \eta  k^2 (\lambda_{1}-D)}{9 a^2 D} \\
    &- \frac{2 V_{X} \left(3 a^2 H^3 \lambda_{6}+H k^2 \lambda_{5}\right)}{9 a D} - \frac{2 H (D+\lambda_{3}) h'}{9 a D} - \frac{H^2 \lambda_{4} V_{X}^{\prime}}{3 D}.
\end{split}
\end{equation}

    \item Anisotropic stress:
\begin{equation}\label{eq:sigmaDE}
\begin{split}
    \left(\rho_{DE}+p_{DE}\right)\sigma_{DE}&= \frac{ \alpha_{\textrm{M}} H }{9 a}\left(6 \eta^{\prime}+h^{\prime}\right) - \frac{  2  k^2\alpha_{\textrm{T}} }{9 a^2}\eta + \frac{2 H k^2  (\alpha_{\textrm{T}}-\alpha_{\textrm{M}})}{9 a}V_{X}\\
    &-\sigma_{m} \left(1-\frac{1}{M^{2}_{*}}\right).
\end{split}
\end{equation}
\end{itemize}
\end{subequations}

From Equations (\ref{eq:drhoDE}-\ref{eq:sigmaDE}) we can see that in the presence of a non-minimally coupled term, $M_* \neq 1$, we have a term coming from the ordinary matter sector in the dark energy fluid quantities. This new contribution follows from our definition of the energy-momentum tensor, in which we chose not to absorb the time-dependent Planck Mass in its definition, in order to have a standard conservation equation for the dark energy energy density. 
%In the case of $G_{4}=G_{5}=0$, this extra contribution vanishes.

Once having evaluated the above quantities, one can implement the relativistic corrections coming from the dark energy perturbations in Equation (\ref{eq:newteqGR}). This is done by adding the extra dark energy fluid contributions in Equations (\ref{eq:HTnbdot}) and (\ref{eq:HTnbdotdot}), as well as computing the additional source term $\delta \rho_{DE}^{\mathrm{Nb}}$. 

In this section we presented a consistent method for introducing dark energy described by Horndeski gravity into Newtonian simulations at the linear level. We are able to do so by considering the new terms coming from Horndeski's theory as an effective fluid. This approach follows from the definition of effective dark energy fluid as a non-interacting fluid, which allows us to use the N-body equations derived in the previous section. In the next section we will move to present an example of our method.

\section{Case study: k-essence}\label{sec:Kess}

\subsection{The model}
In order to show how the steps outlined in the previous section work, we will introduce the relativistic correction coming from a scalar field in the case of k-essence, a subclass of Horndeski's theory~\cite{arm1,arm2}\footnote{In k-essence there is no anisotropic stress.}. K-essence is a natural extension of quintessence models, in which the kinetic term of the scalar field Lagrangian has a non-trivial form, which, in turn, allows the dark energy to cluster above its sound horizon. The density perturbations in such models, however, are suppressed by a term $1+w$ if the dark energy is close to the cosmological constant, $w \to -1$.  

\begin{figure}[htbp!] 
\centering
\includegraphics[width=.48\textwidth]{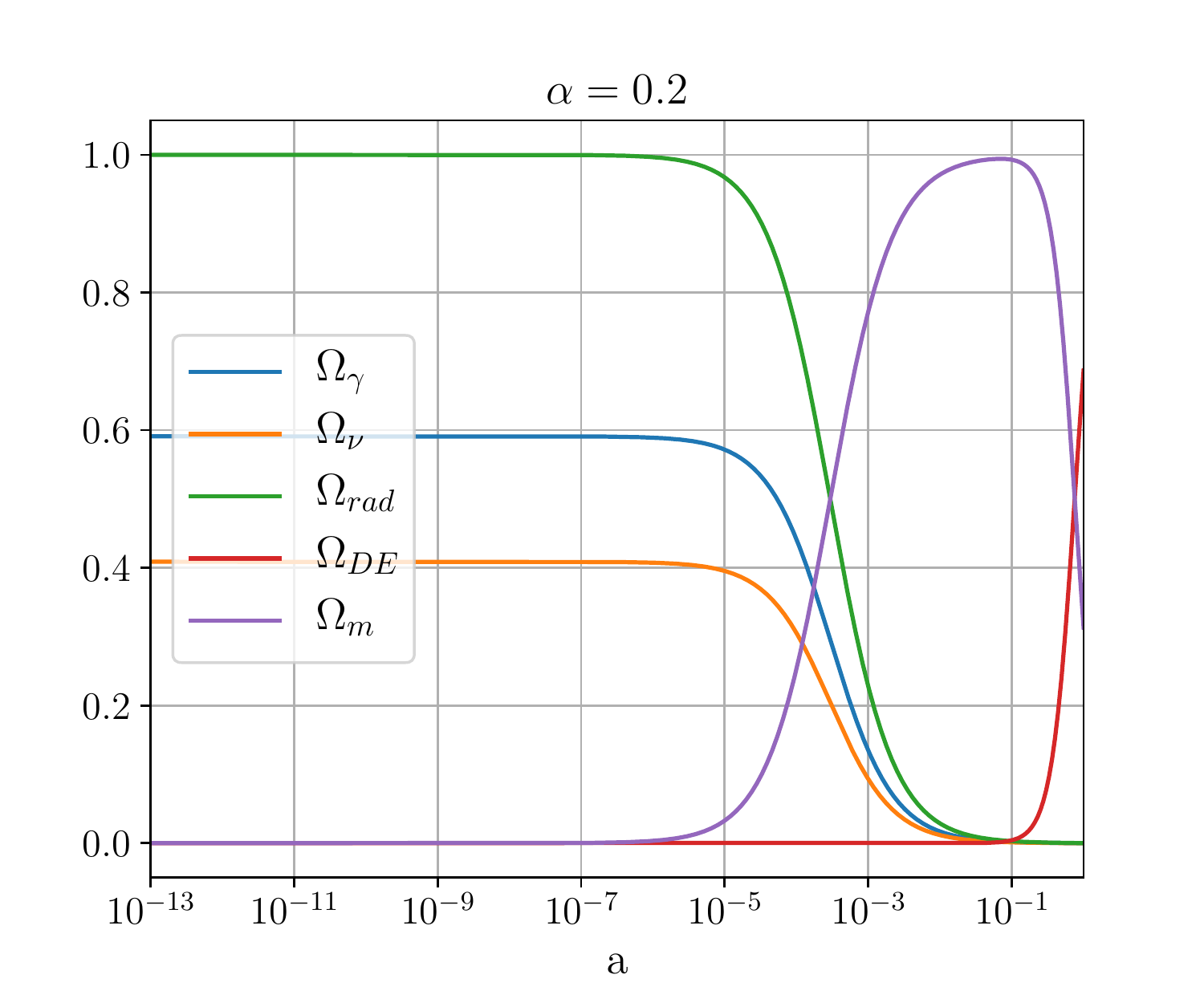} 
\includegraphics[width=.48\textwidth]{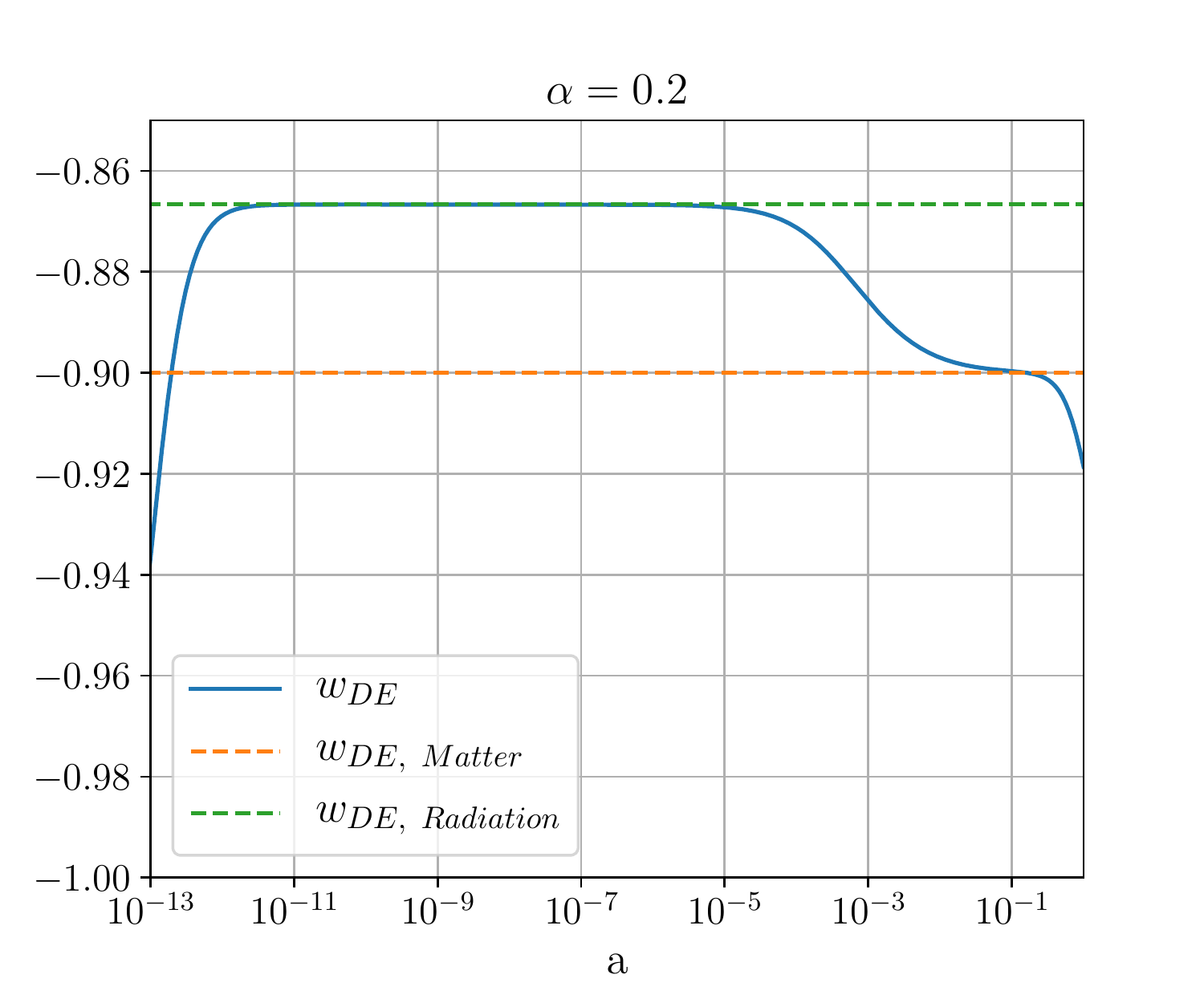}
\caption{Background evolution for the k-essence model given by (\ref{actionkess}), with $\alpha=0.2$ and massless neutrinos. Left plot shows the fractional density of each species as a function of the scale factor. Right plot gives the evolution of the dark energy equation of state as a function of the scale factor, the dashed lines show the constant value of $w_{DE}$ at the two (radiation and matter) domination epochs given by Equation (\ref{eq:wde}). For the scalar field sound speed we can use Equation (\ref{eq:cs2de}), which gives $c_{s}^{2}\sim 0.013$ for $\alpha=0.2$ and during matter domination epoch ($w_{B}=0$).}
\label{fig:Om_w_kess}
\end{figure}

To construct the N-body gauge quantities we introduce k-essence in a fully covariant way in the Einstein-Boltzmann code for Horndeski theories, $\hiclass$. We began with the following action for our implementation~\cite{chibaetal}
\begin{equation}\label{actionkess}
    S = \int \mathrm{d}^{4}x \sqrt{-g} \left[ \frac{1}{2\kappa^{2}}R + p(\phi,X) \right] + S_{M},
\end{equation}
where $\kappa^{2}=8 \pi G$, $S_{M}$ is the matter action for a perfect fluid and the function $p(\phi,X)$ is given by:
\begin{equation}
    p(\phi,X) = \frac{V_{0}}{\phi^{\alpha}}\left(-X+X^{2}\right),
\end{equation}
with $X = -\frac{1}{2}\nabla_{\mu}\phi\nabla^{\mu}\phi$. The action (\ref{actionkess}) has a scaling solution, a desirable feature when setting the initial conditions for the scalar field. We will briefly summarize the properties of such solutions in this model. Scaling solutions are such that the equation of state parameter $w_{\phi}=p_{\phi}/\rho_{\phi}$ remains constant during each era of domination (radiation, matter and dark energy) of the Universe. From (\ref{actionkess}) we have that the pressure and energy density of the scalar field are:
\begin{subequations}
\begin{align}\label{rhopphi}
    &p_{\phi} = p(\phi,X) = \frac{V_{0}}{\phi^{\alpha}}\left( -X + X^{2}\right),\\
    &\rho_{\phi} = 2X\frac{\partial p}{\partial X} - p = \frac{V_{0}}{\phi^{\alpha}}\left(-X+3X^{2}\right).
\end{align}
\end{subequations}
Therefore, when $w_{\phi}$ is constant, $X$ is also constant and can be written as:
\begin{equation}\label{Xscsol}
    X = \frac{1- w_{\phi}}{1-3w_{\phi}}.
\end{equation}
During radiation or matter domination ($\rho_{B}\gg \rho_{\phi}$), where the subscript $B$ refers to the dominant species in the background density, the continuity equation for the scalar field is given by:
\begin{equation}\label{conteqscsol}
    \dot{\rho}_{\phi} = - \frac{2}{t\left(1+w_{B}\right)}\left(1+w_{\phi}\right)\rho_{\phi},
\end{equation}
where a dot denotes the derivative with respect to the cosmic time, $t$. Substituting (\ref{rhopphi}) and (\ref{Xscsol}) into (\ref{conteqscsol}), we have the following relation between the parameter $\alpha$ and the equation of state parameters $w_{\phi}$ and $w_{B}$:
%\begin{equation}
%    \alpha = -2 \frac{\left(1+w_{\phi}\right)}{1+w_{B}},
%\end{equation}
%conversely
\begin{equation}\label{eq:wde}
    w_{\phi} = \frac{\left(1+w_{B}\right)\alpha}{2}-1.
\end{equation}
The sound speed squared of the scalar perturbations can also be written in terms of the extra parameter $\alpha$
\begin{equation}\label{eq:cs2de}
    c_{s}^{2} = \frac{\alpha  (w_{B}+1)}{16-3 \alpha  (w_{B}+1)},
\end{equation}
from which we see that when $\alpha \to 0^{+}$ we have $c_{s}^{2} \to 0$.
Requiring $w_{\phi}<0$ during matter domination, we are left with the condition $\alpha<2$ on the extra parameter of the model. The existence and stability of a dark energy fixed point for this model is carefully presented in~\cite{chibaetal}. We set our initial values for the scalar field as:
\begin{equation}
    \phi_{\text{ini}} = \sqrt{\frac{2\left(1+w_{\phi}\right)}{1-3w_{\phi}}}t
\end{equation}
in the radiation-dominated era. The cosmological parameters used in this work are summarized in Table \ref{table:class_parameters}. We plot the background evolution of this model in Figure \ref{fig:Om_w_kess}.

\begin{table}[tb]
    \begin{center} 
        \begin{tabular}{|l| c| c|} 
            \hline
            Parameter & $\Lambda$CDM  & $\sum m_\nu = 0.10\,\text{eV}$  \\
            \hline
            $A_\text{s}$  & $2.215 \times 10^{-9}$& $2.215 \times 10^{-9}$ \\
            $n_\text{s}$ & $0.9655$ & $0.9655$ \\
            $\tau_\text{reio}$ & $0.078$ & $0.078$  \\
            $\Omega_{\text{b}}$ & $0.049$  & $0.049$  \\
            $\Omega_{\text{cdm}}$ & $0.264$  & $0.262$ \\
            $\Omega_{\nu}$ & $3.77\times 10^{-5}$ & $2.37\times 10^{-3}$  \\  
            $h$ & $0.6731$ & $0.6731$ \\
            $\alpha$ & $0.2$ & $0.2$ \\
            \hline						
        \end{tabular}
    \end{center}
    \caption{Cosmological and K-Essence parameters for the \hiclass \ runs used to generate the plots in this work. We have used the exact relation $\Omega_{\text{cdm}} = 0.2643 - \Omega_{\nu}$.}
    \label{table:class_parameters} 
\end{table}

\subsection{Results}
To obtain the k-essence fluid quantities, we set $D=\alpha_{\textrm{K}}$, $\alpha_{\textrm{B}}=\alpha_{\textrm{M}}=\alpha_{\textrm{T}}=0$ and $M_{*}^{2}=1$ in Equations (\ref{eq:drhoDE}-\ref{eq:sigmaDE}), which leaves us with:
\begin{subequations}
\begin{align}\label{eq:drhokess}
&\delta \rho_{\text{k-ess.}} = -\frac{1}{3} H \left\lbrace a V_X \left[\alpha_{\textrm{K}} H^2 + 9(p_{m}+\rho_{m})\right]+6 V_{X} H^{\prime}-\alpha_{\textrm{K}} H V_{X}^{\prime}\right\rbrace,\\
&\delta p_{\textrm{k-ess.}} = -\frac{2 a H^3 \lambda_{6} }{3 \alpha_{\textrm{K}}}V_{X}-\frac{1}{3} H^2 \lambda_{2}  V_{X}^{\prime},\\
&\theta_{\textrm{k-ess.}} = -k^{2}V_{X}.
\end{align}
\end{subequations}

\begin{figure}[htbp!] 
\centering
\includegraphics[width=0.85\textwidth]{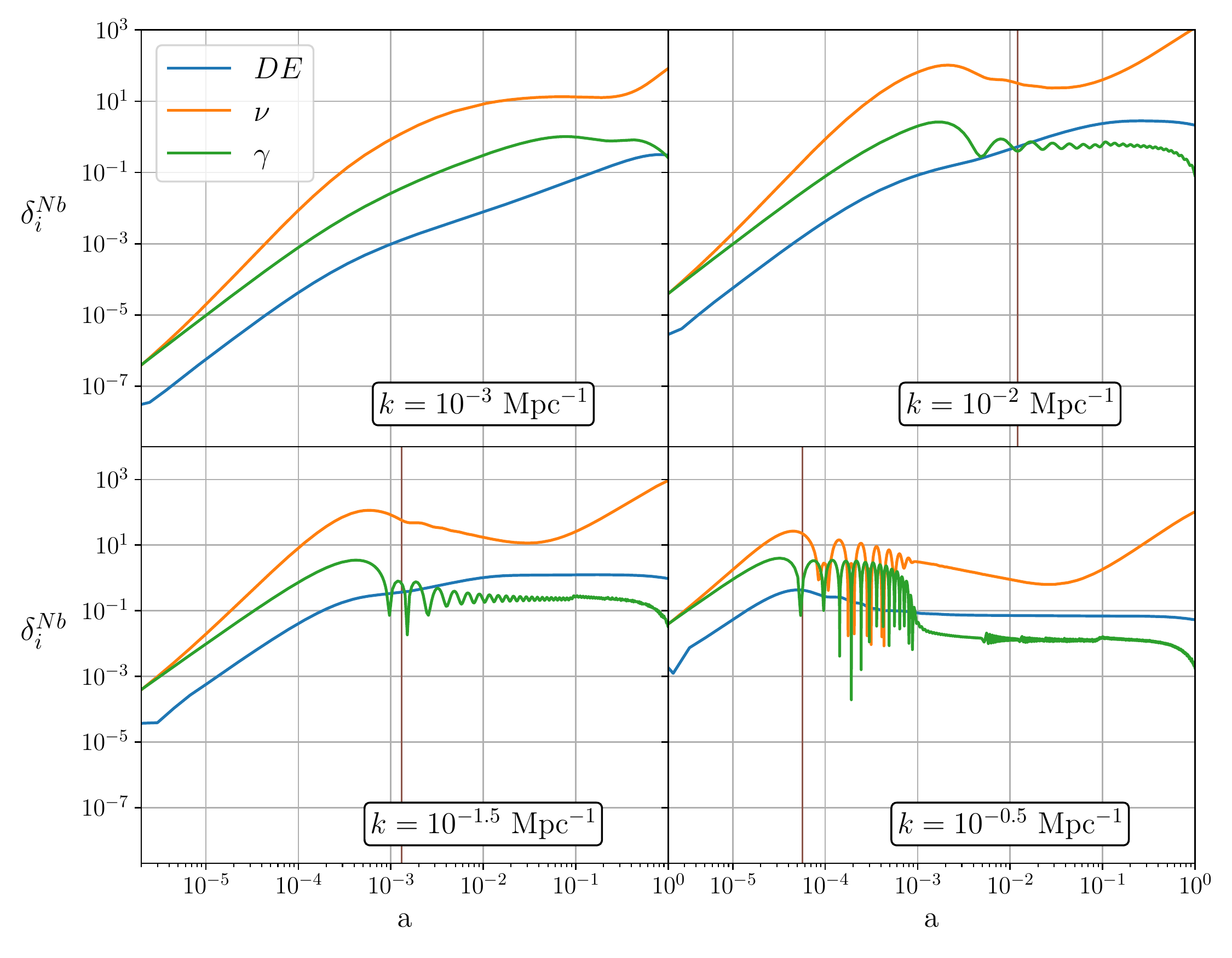} 
\caption{N-body gauge density perturbations as a function of the scale factor for three different species: dark energy, neutrinos and photons, for four different $k$ values. The thin brown vertical lines correspond to the sound horizon of dark energy perturbations, $kc_{s, DE}/ \mathcal{H}=1$, at each $k$ mode. The dark energy perturbation grows above the sound horizon scales. Note that for the first case, the sound horizon crossing happens at $a>1$. The perturbations are normalised so that $\zeta = -1$ on super-horizon scales.}

\label{fig:deltas_nb_diff_ks}
\end{figure}

In Figure \ref{fig:deltas_nb_diff_ks}, we present the time evolution of the N-body gauge density perturbations for dark energy, massive neutrinos with $\sum m_{\nu} =0.1$ eV and photons for four different $k$ values. We also indicate the scale factor when the perturbation enters the sound horizon. The dark energy density perturbations grow before they enter the sound horizon and they freeze in the matter-dominated era inside the sound horizon. The radiation perturbations oscillate and then decay once they enter the horizon. On the other hand, massive neutrinos become non-relativistic for $\sum m_{\nu}=0.1$ eV at 
\begin{equation*}
    z_{nr} = \frac{\sum m_{\nu}}{3.15T_{0,\nu}} - 1 \sim 188,
\end{equation*}
with $T_{0,\nu} \sim 1.9\textrm{K}$ being the temperature of the neutrinos today. The massive neutrino density perturbations under the horizon scale grow like dark matter after this epoch. 

\begin{figure}[htbp!] 
\centering
\includegraphics[width=0.85\textwidth]{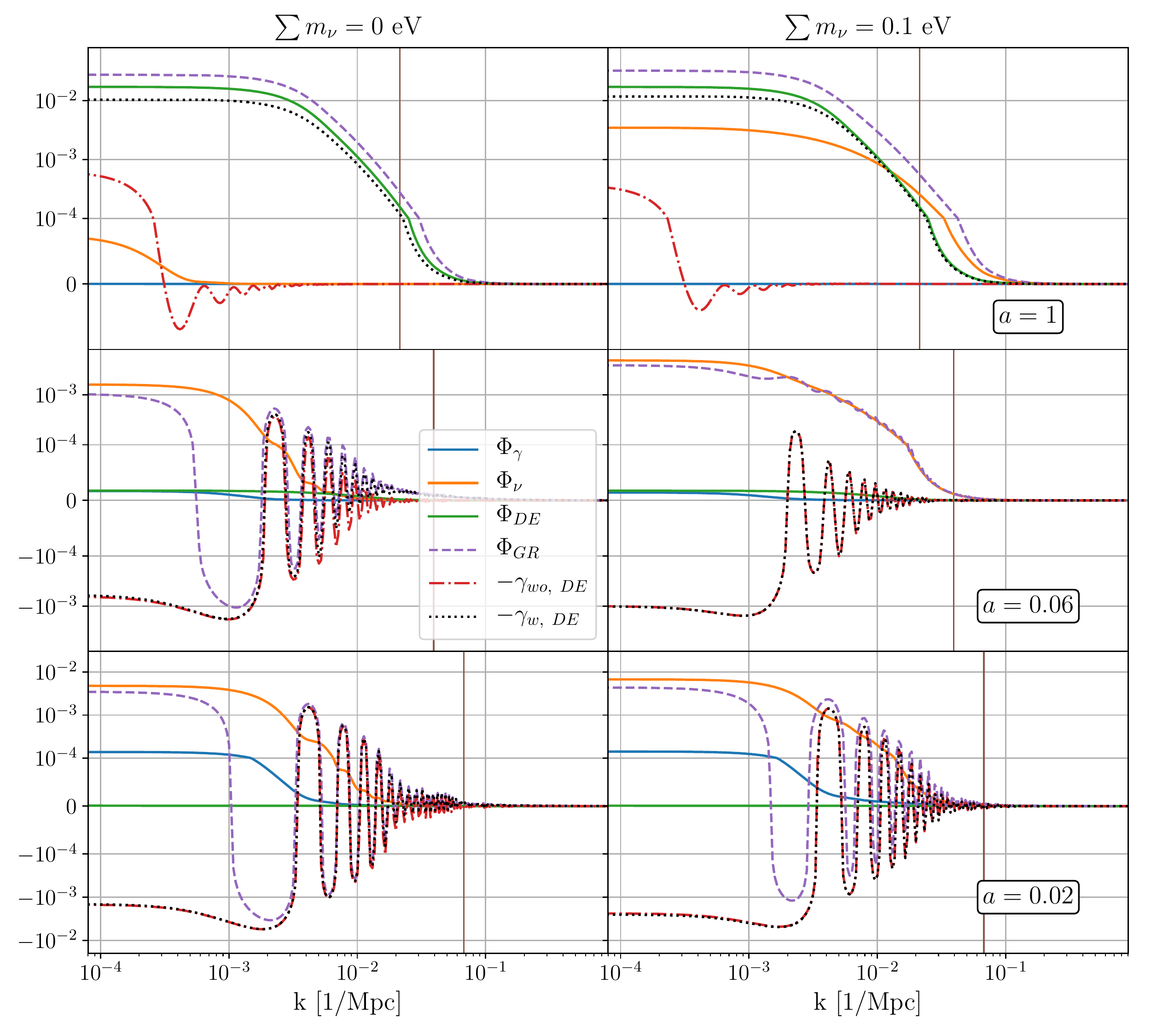} 
\caption{Individual contributions from the ``force'' potentials of each relativistic species, the sum $\Phi_{GR}$ and the relativistic correction potentials with and without dark energy perturbations, $\gammanb_{w, \ DE}$ and $\gammanb_{wo, \ DE}$ respectively. The left plots are for massless neutrinos and the right massive neutrinos, each row is at a given scale factor, $a=1$ (top), $a=0.06$ (middle) and $a=0.02$ (bottom). The thin brown vertical lines correspond to the sound horizon of dark energy perturbations, $k=  \mathcal{H}/c_{s, DE}$. The perturbations are normalised so that $\zeta = -1$ on super-horizon scales.}
\label{fig:phis_gammas_kess}
\end{figure}

We plot the ``force'' potentials of the relativistic species, $k^2\Phi_\alpha=4\pi G a^2 \delta\rho_\alpha^{Nb}$ ($\alpha=\{\gamma,\nu,\textrm{DE}\}$), and the $\gammanb$ contributions in Figure \ref{fig:phis_gammas_kess}. The contribution $\gammanb_{wo, \ DE}$ refers to equation (\ref{eq:k2gammanb}) computed without the dark energy perturbations in it, that is, k-essence is present only in the background quantities. We can see that the density perturbations of dark energy are only relevant at late times, in the $a=1$ (top row) plots, in which the total general relativistic ``force'' potential, $\Phi_{GR}=\Phi_\gamma+\Phi_\nu+\Phi_{DE}+\gammanb$, gets most of its contribution from $\gammanb_{w, \ DE}$. The lack of oscillations for intermediate $k$ values of $\Phi_{DE}$ and $\gammanb_{w, \ DE}$ at redshift $z=0$, stems from the nature of the clustering dark energy density perturbations: the dark energy density grows and the potential $\Phi_{DE}$ remains constant above the sound horizon $c_s/\mathcal{H}$ while it decays below the sound horizon. Thus $\Phi_{DE}$ is non-zero only for $k <  \mathcal{H}/c_{s, DE}$. In contrast, at higher redshifts, since the dominant term in $\Phi_{GR}$ comes from $\gammanb_{wo, \ DE}$, the oscillatory and damped behavior of relativistic species appear. For the massive neutrinos case (right column plots), $\Phi_{\nu}$ does not exhibit any oscillation as the neutrinos have already become non-relativistic. 

\begin{figure}[htbp!] 
\centering
\includegraphics[width=0.85\textwidth]{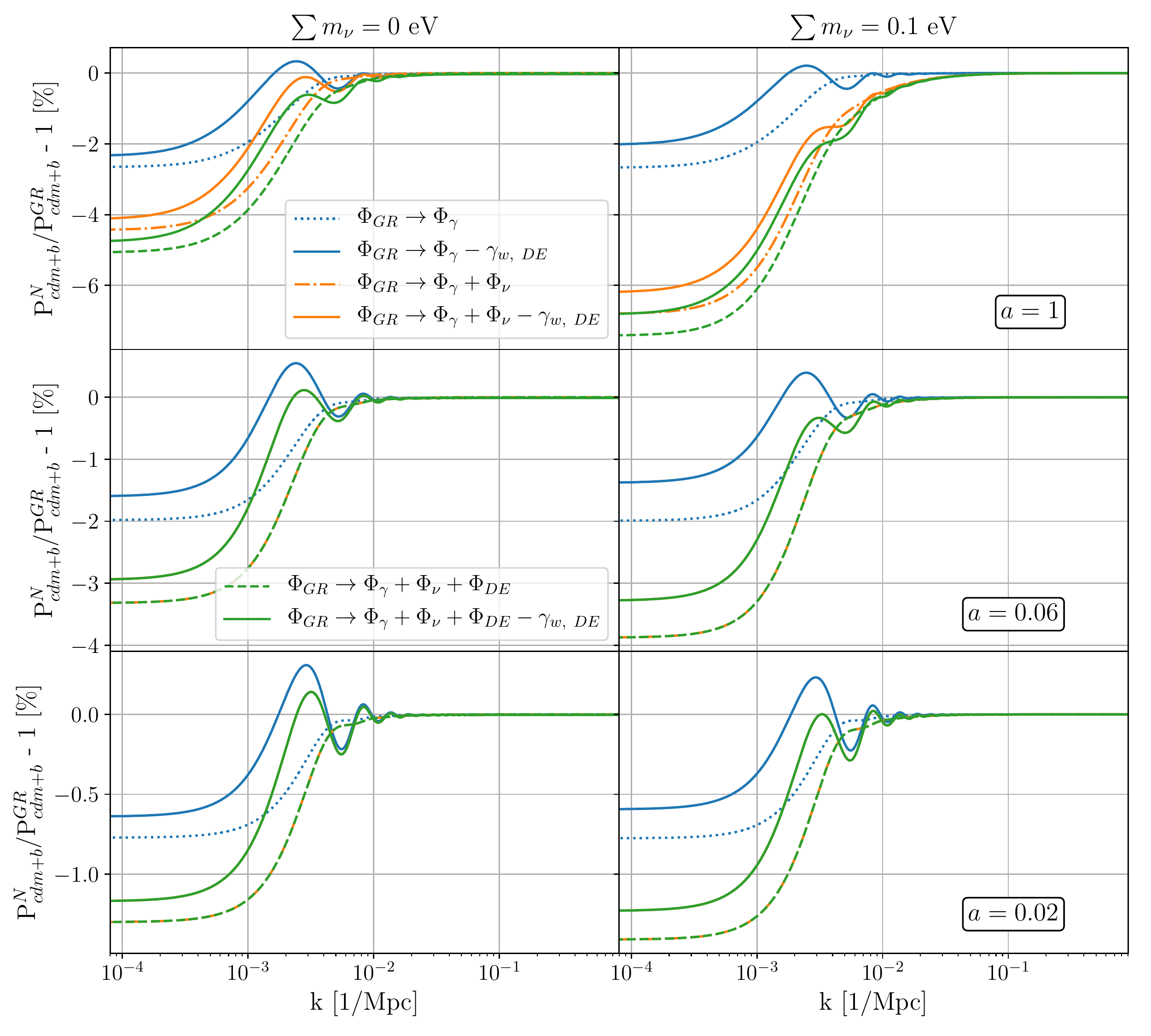} 
\caption{Relative difference to the matter (CDM + baryons) power spectra in the N-Body gauge, with (superscript GR) and without (superscript N) relativistic corrections, at three different scale factors. Left column plots are for the massless neutrinos case, and right for massive neutrinos. Top plots are at $a=1$, middle $a=0.06$ and bottom $a=0.02$. Our initial conditions for $\delta_{cdm+b}^{\textrm{Nb}}$ are set at $a=0.01$ ($z=99$).}
\label{fig:pow_spectra_kess}
\end{figure}

In Figure \ref{fig:pow_spectra_kess} we present the relative matter power spectrum (CDM + baryons) with and without the relativistic correction, computed in the N-body gauge. We see again that at higher redshifts the dark energy ``force'' potential has a negligible effect as expected since the scalar field perturbations are only relevant at smaller redshifts. In both cases of massless and massive neutrinos, k-essence amounts to roughly an additional $1\%$ increase in the relative deviation at $z=0$. Also, the relative difference between the $\Phi_{\gamma}+\Phi_{\nu}$ curves in both cases, are smaller at higher redshifts, as the massive neutrinos are still close to being relativistic. We have compared our results with~\cite{Thomas}, and they are in good agreement, with some small differences arising from the different background evolution as well as the presence of the clustering dark energy component.

%%%%%%%%%%%%%%%%%%%%%%%%%%%%%%%%%%%%%%5

\section{Discussion}\label{sec:Concl}
In the coming years, large areas of the sky will be probed with surveys such as SKA\footnote{\href{www.skatelescope.org}{{\tt www.skatelescope.org}}} and EUCLID\footnote{\href{https://sci.esa.int/web/euclid}{{\tt sci.esa.int/euclid}}}. On the largest scales relativistic effects become important and one needs to properly compute the contribution of components with non-zero pressure, including photons, neutrinos and dark energy.

In this work we have outlined a method to calculate relativistic corrections in Newtonian simulations coming from dark energy in the form of a scalar field described by Horndeski theory. Our approach uses the N-body gauge~\cite{Nbody1}, a specific choice of spacetime coordinates in which there is no volume deformation coming from the metric perturbation, so that the relativistic density is the same as the density computed by counting the number of particles in Newtonian simulations. Our method is valid at linear scales, and for the case of light neutrinos, e.g., mass less than $0.5$~eV that can be treated using linear perturbation theory. 

To consistently introduce the contributions coming from the scalar field in the N-body gauge, we formulated Horndeski's theory using an effective fluid description. We showed how to extract the perturbed fluid quantities ($\delta \rho_{DE}$, $\theta_{DE}$, $\sigma_{DE}$, $\delta p_{DE}$), in terms of the functions $\alpha_{i}$ ($i=\textrm{B},\textrm{M},\textrm{K},\textrm{T}$) that characterise linear perturbations in Horndeski's theoy. These fluid variables can be computed using a linear Einstein-Boltzmann code such as \hiclass~\cite{hiclass1, hiclass2}. Within this framework one can then compute the relativistic correction, $\gammanb$ defined in equation~(\ref{eq:k2gammanb}), including dark energy perturbations, and solve Equation (\ref{eq:newteqGR}) to get the matter (CDM+baryons) density contrast including relativistic effects. The effect of $\gammanb$ can be included in Newtonian N-body simulations, making them consistent with linear relativistic perturbation theory on large scales~\cite{Thomas}. 

In Figure \ref{fig:pow_spectra_kess} we show that the corrections coming from a k-essence scalar field can have a $1\%$ effect in the matter power spectrum of pressureless species, in two scenarios of massless and massive ($\sum m_{\nu}$=0.1 eV) neutrinos. These corrections are relevant at large scales, but are subdominant for small scales. This is expected, since the $1+w_{DE}$ term in the density perturbations for the k-essence model reduces the effect of dark energy clustering, and, therefore, suppresses its contribution to the matter power spectrum. This is a peculiarity of the model that we considered in this paper where the smaller sound speed implies $w_{DE}$ being closer to $-1$. It would be interesting to investigate different k-essence and modified gravity models, in which this suppression is not present in $\delta \rho_{DE}$. Different dark energy models leave different imprints on large scales, and with future $21$-cm surveys we expect that a large enough effective volume survey might make these effects detectable~\cite{mcquinn,mao,loeb,alonso}. 

Finally we comment on the extension of our method to fully non-linear scales. As we mentioned in section 2, the temporal gauge condition that we use in the N-body gauge is liable to break down on non-linear scales. This can be avoided by choosing a different temporal gauge condition such as the one used in the Poisson gauge (as is done in the N-boisson gauge~\cite{nboisson1,nboisson2}). It is also possible to impose an alternative spatial gauge condition, the Newtonian motion gauge, to eliminate all the relativistic corrections in the Euler equation, so that it is equivalent to the non-linear Newtonian equation even in the presence of massive neutrinos and dark energy perturbations \cite{Nbody3}. In this case, relativistic corrections are entirely encoded in metric perturbations and they can be reintroduced to Newtonian simulations by performing a gauge transformation to N-body gauge as a post-processing~\cite{Partmann}. Our method to include modified gravity as a dark energy fluid is readily applicable. 

Our approach has a limitation that dark energy perturbations are treated linearly. This is a good approximation as long as non-linear clustering of the effective dark energy is negligible. In the case of k-essence, this requires that the sound speed of the scalar field is not too small \cite{farbod1}. The modified gravity parameters $\alpha_i$ are strongly constrained by various observations and the linear approximation is expected to work well in general. For example, $\alpha_{\textrm{M}}$, $\alpha_{\textrm{B}}$ and $\alpha_{\textrm{T}}$ are strongly constrained by the solar system tests ($\alpha_{\textrm{M}} <0.002$ for shift-symmetric theories \cite{Burrage:2020jkj}) and gravitational wave observations ($\alpha_{\textrm{B}} < 0.01$ from gravitational wave instabilities \cite{GWinstability} and $\alpha_{\textrm{T}} < 10^{-15}$ from the speed of gravitational waves). On the other hand, $\alpha_{\textrm{K}}$ is relatively unconstrained as this parameter does not affect the perturbations under the quasi-static approximation and our approach is ideal to include the effect of $\alpha_{\textrm{K}}$ in Newtonian simulations as we did for k-essence models. However, we should note that the linear approximation for the scalar field breaks down in some theories on small scales. This can be seen from the field equation for the scalar field perturbations $V_X$ defined in equation~(\ref{def:VX}):
\begin{align}
D\left(2-\alpha_{\textrm{B}}\right)V_{X}^{\prime\prime}+8aH\lambda_{7}V_{X}^{\prime} & +2a^{2}H^{2}\left[\frac{c_{\text{sN}}^{2}k^{2}}{a^{2}H^{2}}-4\lambda_{8}\right]V_{X}=\frac{2c_{\text{sN}}^{2}}{aH}k^{2}\eta\label{eq:metric_vx}\nonumber \\
 & +\frac{3a}{2HM_{*}^{2}}\left[2\lambda_{2}\delta\rho_{\textrm{m}}-3\alpha_{\textrm{B}}\left(2-\alpha_{\textrm{B}}\right)\delta p_{\textrm{m}}\right]\,, 
\end{align}
where $c_{\text{sN}}^{2}$ in the numerator is the sound speed squared of the scalar field, which is defined Appendix~\ref{sec:AppA} along with the functions $D$ and $\lambda_i$.
If $\lambda_2 \neq 0$, the scalar field perturbation is sourced by the matter density perturbation, which becomes non-linear on small scales. We then need to take into account non-linear corrections to the equations of motion. This is relevant to the models with screening mechanisms that rely on the non-linearity of the scalar field perturbations to restore general relativity on small scales. N-body simulations have been developed to deal with these theories \cite{MGNbody} by using the quasi-static approximation and keeping only terms relevant in the large $k$ limit. Our method can be used to make these simulations fully relativistic by including corrections that are missing on large scales.

%%%%%%%%%%%%%%%%%%%%%%
\acknowledgments
We thank Christian Fidler, Thomas Tram and Miguel Zumalac{\'a}rregui for useful discussions. 
GB acknowledges support from the State Scientific and Innovation Funding Agency of Esp\'irito Santo (FAPES, Brazil) and the Coordena\c{c}\~ao de Aperfei\c{c}oamento de Pessoal de N\'ivel Superior - Brasil (CAPES) - Finance Code 001. KK has received funding from the European Research Council
(ERC) under the European Union’s Horizon 2020 research and innovation programme (grant agreement No. 646702 ``CosTesGrav”). KK and DW are supported by the UK Science and Technologies Facilities Council grants ST/S000550/1. 

%%%%%%%%%%%%%%%%%%%%%%
\appendix 

\section{$\alpha$ and $\lambda$ functions}\label{sec:AppA}

In this appendix we present the definitions of the functions $\alpha_{i}$ ($i= \textrm{B, M, K, T}$) and the $\lambda_{i}$ functions , ($i=1,...,8$), shown in Section~\ref{sec:Hornfluid}. These are defined in~\cite{hiclass1}.

\begin{align}
M_{*}^{2}\equiv & 2\left(G_{4}-2XG_{4X}-\frac{H\phi^{\prime}XG_{5X}}{a}+XG_{5\phi}\right)\\
\alpha_{\textrm{M}}\equiv & \frac{d\ln M_{*}^{2}}{d\ln a}\\
H^{2}M_{*}^{2}\alpha_{\textrm{K}}\equiv & 2X\left(G_{2X}+2XG_{2XX}-2G_{3\phi}-2XG_{3\phi X}\right)\\
 & +\frac{12H\phi^{\prime}X}{a}\left(G_{3X}+XG_{3XX}-3G_{4\phi X}-2XG_{4\phi XX}\right)\nonumber \\
 & +12H^{2}X\left[G_{4X}-G_{5\phi}+X\left(8G_{4XX}-5G_{5\phi X}\right)+2X^{2}\left(2G_{4XXX}-G_{5\phi XX}\right)\right]\nonumber \\
 & +\frac{4H^{3}\phi^{\prime}X}{a}\left(3G_{5X}+7XG_{5XX}+2X^{2}G_{5XXX}\right)\nonumber \\
HM_{*}^{2}\alpha_{\textrm{B}}\equiv & \frac{2\phi^{\prime}}{a}\left(XG_{3X}-G_{4\phi}-2XG_{4\phi X}\right)+8HX\left(G_{4X}+2XG_{4XX}-G_{5\phi}-XG_{5\phi X}\right) \label{eq:aB}\\
 & +\frac{2H^{2}\phi^{\prime}X}{a}\left(3G_{5X}+2XG_{5XX}\right)\nonumber \\
M_{*}^{2}\alpha_{\textrm{T}}\equiv & 4X\left(G_{4X}-G_{5\phi}\right)-\frac{2}{a^{2}}\left(\phi^{\prime\prime}-2aH\phi^{\prime}\right)XG_{5X}\,.
\end{align}
As first mentioned in~\cite{belsaw}, each of these functions is independent of the others, and they describe different physical effects individually.

The $\lambda_{i}$ functions are:

\begin{align}
D= & \alpha_{\textrm{K}}+\frac{3}{2}\alpha_{\textrm{B}}^{2}\\
\lambda_{1}= & \alpha_{\textrm{K}}\left(1+\alpha_{\textrm{T}}\right)-3\alpha_{\textrm{B}}\left(\alpha_{\textrm{M}}-\alpha_{\textrm{T}}\right)\\
\lambda_{2}= & -\frac{3\left(\rho_{\textrm{m}}+p_{\textrm{m}}\right)}{H^{2}M_{*}^{2}}-\left(2-\alpha_{\textrm{B}}\right)\frac{H^{\prime}}{aH^{2}}+\frac{\alpha_{\textrm{B}}^{\prime}}{aH}\\
\lambda_{3}= & -\frac{1}{2}\left(2+\alpha_{\textrm{M}}\right)D-\frac{3}{4}\alpha_{\textrm{B}}\lambda_{2}\\
\lambda_{4}= & \alpha_{\textrm{K}}\lambda_{2}-\frac{2\alpha_{\textrm{K}}\alpha_{\textrm{B}}^{\prime}-\alpha_{\textrm{B}}\alpha_{\textrm{K}}^{\prime}}{aH}\\
\lambda_{5}= & \frac{3}{2}\alpha_{\textrm{B}}^{2}\left(1+\alpha_{\textrm{T}}\right)+\left(D+3\alpha_{\textrm{B}}\right)\left(\alpha_{\textrm{M}}-\alpha_{\textrm{T}}\right)+\frac{3}{2}\alpha_{\textrm{B}}\lambda_{2}\\
\lambda_{6}= & \left(1-\frac{3\alpha_{\textrm{B}}H^{\prime}}{\alpha_{\textrm{K}}aH^{2}}\right)\frac{\alpha_{\textrm{K}}\lambda_{2}}{2}-\frac{DH^{\prime}}{aH^{2}}\left[2+\alpha_{\textrm{M}}+\frac{H^{\prime\prime}}{aHH^{\prime}}\right]-\frac{2\alpha_{\textrm{K}}\alpha_{\textrm{B}}^{\prime}-\alpha_{\textrm{B}}\alpha_{\textrm{K}}^{\prime}}{2aH}-\frac{3\alpha_{\textrm{K}}p_{\textrm{m}}^{\prime}}{2aH^{3}M_{*}^{2}}\\
\lambda_{7}= & \frac{D}{8}\left(2-\alpha_{\textrm{B}}\right)\left[4+\alpha_{\textrm{M}}+\frac{2H^{\prime}}{aH^{2}}+\frac{D^{\prime}}{aHD}\right]+\frac{D}{8}\lambda_{2}\\
\lambda_{8}= & -\frac{\lambda_{2}}{8}\left(D-3\lambda_{2}+\frac{3\alpha_{\textrm{B}}^{\prime}}{aH}\right)+\frac{1}{8}\left(2-\alpha_{\textrm{B}}\right)\left[\left(3\lambda_{2}-D\right)\frac{H^{\prime}}{aH^{2}}-\frac{9\alpha_{\textrm{B}}p_{\textrm{m}}^{\prime}}{2aH^{3}M_{*}^{2}}\right]\label{eq:lambda_8}\\
 & -\frac{D}{8}\left(2-\alpha_{\textrm{B}}\right)\left[4+\alpha_{\textrm{M}}+\frac{2H^{\prime}}{aH^{2}}+\frac{D^{\prime}}{aHD}\right]\nonumber \\
c_{\text{sN}}^{2}= & \lambda_{2}+\frac{1}{2}\left(2-\alpha_{\textrm{B}}\right)\left[\alpha_{\textrm{B}}\left(1+\alpha_{\textrm{T}}\right)+2\left(\alpha_{\textrm{M}}-\alpha_{\textrm{T}}\right)\right]\,.
\end{align}
Where $c_{\text{sN}}^{2}$ is the numerator of the sound speed squared of the
scalar field
\begin{equation}
\label{def:csN}
c_{\text{s}}^{2}=\frac{c_{\text{sN}}^{2}}{D}\,.
\end{equation}

%%%%%%%%%%%%%%%%%%%%%%%%%%%%%%%%% 

\end{document}